\documentclass{article}
\usepackage{arxiv}
\usepackage[utf8]{inputenc} 
\usepackage[T1]{fontenc}    
\usepackage{hyperref}       
\usepackage{url}            
\usepackage{booktabs}       
\usepackage{amsfonts}       
\usepackage{nicefrac}       
\usepackage{microtype}      
\usepackage{graphicx}
\usepackage{natbib}
\usepackage{doi}
\usepackage{listings}
\usepackage{xcolor}
\usepackage[margin=1in]{geometry} 

\definecolor{codegreen}{rgb}{0,0.6,0}
\definecolor{codegray}{rgb}{0.5,0.5,0.5}
\definecolor{codepurple}{rgb}{0.58,0,0.82}
\definecolor{backcolour}{rgb}{0.95,0.95,0.92}

\lstdefinestyle{mystyle}{
    backgroundcolor=\color{backcolour},   
    commentstyle=\color{codegreen},
    keywordstyle=\color{magenta},
    numberstyle=\tiny\color{codegray},
    stringstyle=\color{codepurple},
    basicstyle=\ttfamily\footnotesize,
    breakatwhitespace=false,         
    breaklines=true,                 
    captionpos=b,                    
    keepspaces=true,                 
    numbers=left,                    
    numbersep=5pt,                   
    showspaces=false,                
    showstringspaces=false,
    showtabs=false,                  
    tabsize=2
}

\title{Leveraging Artificial Intelligence as a Strategic Growth Catalyst for Small and Medium-sized Enterprises}

\author{
  Oluwatosin Agbaakin \\
  Indiana University \\
  \texttt{\href{mailto:olbaagba@iu.edu}{olbaagba@iu.edu}} \\
}

\date{\today}

\hypersetup{
pdftitle={Leveraging Artificial Intelligence as a Strategic Growth Catalyst for Small and Medium-sized Enterprises},
pdfsubject={cs.AI, econ.GN},
pdfauthor={Oluwatosin Agbaakin},
pdfkeywords={Artificial Intelligence, Machine Learning, SME, Business Growth, Entrepreneurship, Strategic Framework, Knowledge Graphs},
}

\begin{document}
\maketitle

\begin{abstract}
Artificial Intelligence (AI) has transitioned from a futuristic concept reserved for large corporations to a present-day, accessible, and essential growth lever for Small and Medium-sized Enterprises (SMEs). For entrepreneurs and business leaders, strategic AI adoption is no longer an option but an imperative for competitiveness, operational efficiency, and long-term survival. This report provides a comprehensive framework for SME leaders to navigate this technological shift, offering the foundational knowledge, business case, practical applications, and strategic guidance necessary to harness the power of AI. The quantitative evidence supporting AI adoption is compelling; 91\% of SMEs using AI report that it directly boosts their revenue \cite{salesforce2024smb}. Beyond top-line growth, AI drives profound operational efficiencies, with studies showing it can reduce operational costs by up to 30\% and save businesses more than 20 hours of valuable time each month \cite{superagi2024, colorwhistle2025}. This transformation is occurring within the context of a seismic economic shift; the global AI market is projected to surge from \$233.46 billion in 2024 to an astonishing \$1.77 trillion by 2032 \cite{fortunebiz2024}. This paper demystifies the core concepts of AI, presents a business case based on market data, details practical applications, and lays out a phased, actionable adoption strategy.
\end{abstract}


\section{Demystifying the Digital Workforce: AI and Machine Learning in a Business Context}

To strategically deploy AI, business leaders must first understand its core components not as abstract technical terms, but as practical business tools. This section decodes the essential terminology, providing a clear framework for thinking about how these technologies create value.

\subsection{From Buzzwords to Business Tools}

The terms Artificial Intelligence, Machine Learning, and algorithms are often used interchangeably, but they represent distinct layers of a powerful technological stack.

\begin{itemize}
    \item \textbf{Artificial Intelligence (AI)} is the broadest concept, referring to the general ability of computers to emulate human cognitive functions like problem-solving, reasoning, and learning \cite{redhat2024, columbia2024, ibm2024business}. In a business context, AI is the overall field dedicated to creating systems that can perform tasks that typically require human intelligence, from understanding customer queries to identifying market trends. It is best understood as the entire discipline, much like "transportation" is the discipline of moving things from one place to another.

    \item \textbf{Machine Learning (ML)} is the most prevalent and practical application of AI today. It is a subset of AI that uses algorithms to analyze large amounts of data, learn from the patterns within it, and then make decisions or predictions without being explicitly programmed for every possible scenario \cite{redhat2024, googlecloud2024ai, ibm2024ml}. ML is the engine that powers most modern AI tools. Continuing the transportation analogy, if AI is the car, then machine learning is the internal combustion engine—the specific technology that consumes fuel (data) to generate power (predictions and insights).

    \item \textbf{Algorithms} are the specific, step-by-step mathematical procedures or sets of rules that a machine learning system uses to process data, learn, and arrive at a decision \cite{ibm2024ml, sba2025}. They are the detailed blueprints that define how the engine works, dictating the precise sequence of operations required to turn data into a useful output.

    \item \textbf{Deep Learning} represents the cutting edge of machine learning. It is an advanced method that uses complex structures called neural networks—inspired by the human brain—to learn from vast quantities of unstructured data like text, images, and audio with minimal human supervision \cite{redhat2024, columbia2024, ibm2024business}. This is the technology behind the recent explosion in generative AI tools like ChatGPT. In our analogy, deep learning is a high-performance, self-tuning racing engine capable of processing complex fuel mixtures (unstructured data) to achieve superior performance.
\end{itemize}

The fundamental shift for business leaders to grasp is the move from a world of \textit{explicit programming} to one of \textit{probabilistic training}. Traditional software required developers to write precise `if-then` rules to cover every conceivable situation, a process that is rigid and cannot scale to handle the complexity of real-world business problems like predicting customer behavior \cite{googlecloud2024ai}. Machine learning, in contrast, learns these rules implicitly by analyzing historical data \cite{redhat2024, ibm2024ml}. This means the quality of an AI system's output is directly and inextricably linked to the quality and quantity of the data it is fed. Consequently, the most valuable asset for an SME in the AI era is not its software code but its unique, proprietary data. A local retailer's sales history, a consulting firm's project notes, or a manufacturer's equipment sensor logs become the high-octane fuel for their custom AI engine. This elevates the strategic importance of actively collecting, cleaning, and unifying business data from a secondary IT task to a primary driver of competitive advantage \cite{synergy2024}.

\subsection{The Machine Learning Toolkit for SMEs}

Machine learning is not a monolithic technology; it is a collection of techniques, each suited for different business problems. For SMEs, three main types are particularly relevant:

\begin{itemize}
    \item \textbf{Supervised Learning:} This is the most common and straightforward type of ML. The model is trained on a dataset where the "right answers" are already known (labeled data) \cite{ibm2024ml}. The algorithm learns the relationship between the inputs and the corresponding outputs and can then predict the output for new, unseen inputs.
    \begin{itemize}
        \item \textbf{Business Application:} Predicting which customers are at high risk of churning. The model is trained on historical customer data, where each customer is labeled as either "churned" or "not churned." The trained model can then analyze current customers and assign a churn-risk score to each.
    \end{itemize}

    \item \textbf{Unsupervised Learning:} In this approach, the model is given unlabeled data and tasked with finding hidden patterns, structures, or groupings on its own \cite{ibm2024ml}. It is used for exploratory data analysis and discovering insights that are not immediately obvious.
    \begin{itemize}
        \item \textbf{Business Application:} Customer segmentation. An e-commerce business can feed its entire customer purchase history into an unsupervised clustering algorithm. The algorithm might automatically identify distinct segments like "high-value bargain hunters," "seasonal shoppers," and "brand loyalists" without any prior definitions, allowing for highly targeted marketing campaigns.
    \end{itemize}

    \item \textbf{Reinforcement Learning:} This type of learning trains a model to make a sequence of decisions in a dynamic environment to maximize a cumulative reward \cite{ibm2024ml}. The model learns through trial and error, receiving positive or negative feedback for its actions.
    \begin{itemize}
        \item \textbf{Business Application:} Dynamic pricing. An online store could use a reinforcement learning agent to adjust the price of a product in real-time. The agent's "reward" is total revenue. It learns to lower the price to stimulate demand during slow periods and raise it during high-demand surges, automatically finding the optimal balance to maximize profit.
    \end{itemize}
\end{itemize}

\section{The Unignorable Business Case: Market Dynamics and the AI Imperative}

The adoption of AI is not merely a technological trend; it is a fundamental economic transformation reshaping industries on a global scale. For SMEs, understanding the magnitude of this shift is the first step toward recognizing AI as a strategic necessity rather than a discretionary investment.

\subsection{A Tectonic Shift in the Global Economy}

The economic momentum behind AI is staggering. Market analyses project the global AI market will experience explosive growth, expanding from a valuation of \$233.46 billion in 2024 to an estimated \$1,771.62 billion by 2032. This trajectory represents a compound annual growth rate (CAGR) of 29.20\%, signaling one of the most rapid technological expansions in modern history \cite{fortunebiz2024, marketsandmarkets2025}.

\begin{figure}[htbp]
  \centering
  \includegraphics[width=0.8\textwidth]{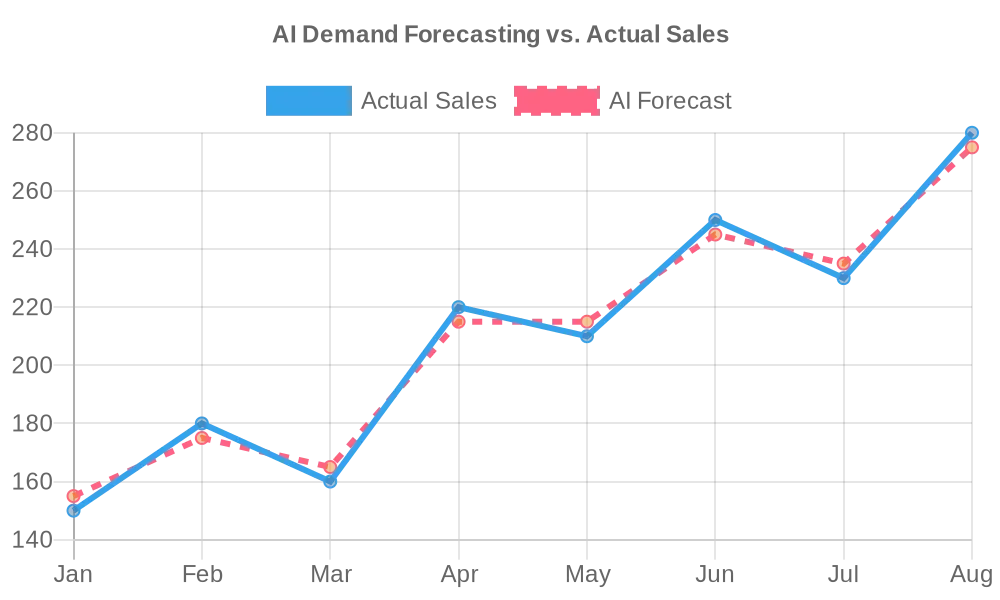}
  \caption{Projected Growth of the Global Artificial Intelligence Market, 2024-2032. Data sourced from Fortune Business Insights and MarketsandMarkets \cite{fortunebiz2024, marketsandmarkets2025}.}
  \label{fig:fig1}
\end{figure}

This market growth is a reflection of AI's profound impact on productivity and economic value. Projections indicate that AI could contribute as much as \textbf{\$15.7 trillion} to the global economy by 2030 \cite{explodingtopics2025}. Research from IDC further quantifies this impact, predicting a cumulative global economic value of \$22.3 trillion by 2030. Their analysis reveals a powerful multiplier effect: every new dollar spent on AI solutions and services by an adopting company is expected to generate an additional \$4.90 in the broader global economy, highlighting AI's role as a catalyst for widespread business acceleration \cite{microsoft2025idc}.

\subsection{The SME Adoption Landscape: An Urgent Call to Action}

While large corporations have been early adopters, SMEs are increasingly recognizing AI's potential. Globally, 77\% of small businesses now report using AI tools in at least one business function, such as customer service or marketing \cite{colorwhistle2025}. However, a significant and potentially dangerous adoption gap persists. A 2025 survey by the World Trade Organization and the International Chamber of Commerce found that \textbf{only 41\% of small firms use AI, compared to over 60\% of large firms} \cite{wto2025}.

More telling is the divide emerging within the SME sector itself. A recent Salesforce survey reveals that growing SMBs are the primary drivers of adoption, with \textbf{83\% of them already experimenting with AI}. Furthermore, 78\% of these growing businesses plan to increase their AI investments in the coming year \cite{salesforce2024smb}. This proactive stance contrasts sharply with stagnant or declining businesses, creating a widening chasm where AI-native SMEs are poised to pull away from their less technologically advanced competitors.

This data paints a clear picture: AI is transitioning from a niche advantage to a foundational business utility, akin to the adoption of electricity or the internet in previous eras. Early adopters of electricity did not just replace their gas lamps; they fundamentally redesigned their factories to leverage the new power source, enabling innovations like the assembly line. Similarly, the greatest benefits from AI are realized not by simply plugging in a new tool, but by redesigning core business workflows around its capabilities \cite{mckinsey2025}. The strong correlation between business growth and AI adoption suggests that AI is becoming a key ingredient \textit{for} growth, not just a byproduct of it. For an SME, the strategic question is no longer "Should we use AI?" but "How do we rewire our business to run on AI?" Answering this question is critical for survival and growth in the coming decade.

\section{The AI-Powered Growth Engine: A Functional Deep Dive}

Artificial intelligence provides SMEs with a suite of powerful tools to enhance performance across every facet of the organization. For lean teams operating with limited resources, AI acts as a "force multiplier," enabling a small group of employees to achieve the analytical depth and operational capacity of a much larger enterprise. This directly addresses the resource constraints that have historically limited SME growth, allowing them to compete more effectively \cite{aws2025, tandfonline2024}.

\subsection{Revolutionizing Marketing and Sales}

AI is transforming marketing from a function based on broad assumptions to a science of precision engagement.

\begin{itemize}
    \item \textbf{Hyper-Personalization at Scale:} Machine learning algorithms can analyze customer browsing history, purchase patterns, and demographic data to deliver uniquely tailored experiences. This moves beyond simple name personalization in emails to recommending specific products, suggesting relevant content, and offering timely promotions that resonate with individual needs \cite{colorwhistle2025, smartosc2024}. For example, an online craft store could use an unsupervised clustering algorithm to automatically identify a "weekend DIYer" segment and schedule marketing emails with project ideas to be sent specifically on Friday afternoons.
    \item \textbf{Predictive Lead Scoring \& AI-Powered CRM:} Instead of treating all leads equally, AI-powered Customer Relationship Management (CRM) systems can analyze lead characteristics and behaviors to assign a "propensity to buy" score. This allows sales teams to focus their limited time on the most promising prospects, dramatically increasing efficiency. Studies show that AI in sales can increase leads by 50\% and reduce call times by 60\% \cite{explodingtopics2025}. Leading tools in this space include \textbf{Salesforce Einstein} and \textbf{HubSpot AI} \cite{salesforce2024crm, hubspot2024crm, kipwise2024}.
    \item \textbf{Automated Content Creation:} Generative AI has become a game-changer for small marketing teams. These tools can rapidly draft blog posts, social media updates, email newsletters, and ad copy, overcoming writer's block and significantly accelerating content production schedules \cite{colorwhistle2025, makebot2025, advansappz2024}. Popular and accessible tools include \textbf{Jasper}, \textbf{ChatGPT}, and \textbf{Google Gemini} \cite{sproutsocial2024, webwave2025, nextiva2025, kipwise2024, workspace2024}.
\end{itemize}

\subsection{Transforming the Customer Experience}

Exceptional customer service is a key differentiator for SMEs. AI enables them to provide responsive, intelligent support at a scale that was previously unimaginable.

\begin{itemize}
    \item \textbf{24/7 Intelligent Support:} Modern AI-powered chatbots can handle a significant portion of routine customer inquiries—reports suggest up to 80\%—without human intervention \cite{colorwhistle2025}. They can answer frequently asked questions, track order statuses, and book appointments around the clock. This frees up human agents to focus on more complex, high-value customer issues. For 72\% of small businesses using AI-driven support, this results in faster resolution times and higher customer satisfaction \cite{colorwhistle2025}. Accessible chatbot platforms include \textbf{Tidio}, \textbf{Nextiva}, and \textbf{Zoho's Zobot} \cite{sproutsocial2024, nextiva2025, fultonbank2024}.
    \item \textbf{Customer Sentiment Analysis:} Using Natural Language Processing (NLP), AI tools can automatically analyze customer communications—such as emails, product reviews, and social media comments—to determine the underlying emotional tone (positive, negative, or neutral). This provides a real-time pulse on customer satisfaction, acting as an early warning system for widespread issues and highlighting brand advocates.
\end{itemize}

\subsection{Optimizing Operations and Supply Chain}

AI brings a new level of predictive power and efficiency to the core operations of a business.

\begin{itemize}
    \item \textbf{Intelligent Demand Forecasting:} By analyzing historical sales data, seasonality, weather patterns, and market trends, ML regression models can predict future product demand with far greater accuracy than traditional methods. This allows SMEs to optimize inventory levels, preventing costly stockouts of popular items and reducing capital tied up in slow-moving overstock \cite{smartosc2024, activdev2025churn}. A small fashion retailer, for instance, could use a time-series forecasting model to more accurately predict demand for winter coats, leading to a 25\% reduction in end-of-season markdowns.
    \item \textbf{Predictive Maintenance:} For SMEs in manufacturing or those reliant on critical machinery, AI is invaluable. Sensors can collect operational data from equipment (e.g., temperature, vibration), and ML models can analyze this data to predict when a part is likely to fail. This allows for maintenance to be scheduled proactively, preventing catastrophic failures and costly operational downtime \cite{redhat2024, mckinsey2019}.
    \item \textbf{Route Optimization:} For businesses involved in delivery or logistics, AI algorithms can analyze traffic data, delivery locations, and vehicle capacity to calculate the most efficient routes in real time. This leads to significant savings in fuel costs and employee time, while also improving delivery speed and customer satisfaction \cite{marketsandmarkets2025, smartosc2024}.
\end{itemize}

\subsection{Streamlining Finance and Administration}

AI automates and enhances the accuracy of back-office functions, reducing administrative burden and providing deeper financial insights.

\begin{itemize}
    \item \textbf{Automated Expense Management and Bookkeeping:} AI-powered finance tools can use computer vision to scan receipts, extract relevant information (vendor, date, amount), automatically categorize expenses, and reconcile transactions with bank statements. This drastically reduces manual data entry, minimizes human error, and can cut accounting costs by up to 50\% \cite{superagi2024}. Leading platforms in this area include \textbf{Docyt} and \textbf{Vic.ai} \cite{gusto2024}.
    \item \textbf{Fraud Detection:} Machine learning models excel at identifying anomalies in patterns. In finance, they can monitor millions of transactions in real-time to flag behavior that deviates from the norm, such as an unusually large purchase or a transaction from a new geographic location, thereby preventing fraudulent activity before significant losses occur \cite{redhat2024, smartosc2024}.
    \item \textbf{Intelligent Financial Forecasting:} Beyond simple bookkeeping, AI can analyze historical cash flow, revenue streams, and external market data to generate more accurate and dynamic financial forecasts. This provides business leaders with a clearer view of future financial health, enabling more strategic decisions about investment, hiring, and expansion \cite{gusto2024, googlecloud2024finance}.
\end{itemize}

\section{Visualizing Success: Using Graphs to Monitor AI-Driven Insights}

While AI models provide powerful predictions, their greatest strategic value for SMEs is unlocked when they move beyond being a "black box" and provide interpretable insights. The true goal is not just to know \textit{what} will happen, but to understand \textit{why}. This is where the synergy between AI and graph-based data visualization becomes critical for tracking relationships and monitoring trends.

Business data is inherently relational. A customer \textit{buys} a product, a marketing campaign \textit{targets} a demographic, and a supplier \textit{provides} a component. As argued by \citet{abdulrahman2024graphs}, representing this information in a relational structure, such as a graph, is fundamental to building a true understanding of the complex system that is a business. An AI model can analyze this relational data to construct a dynamic \textbf{business knowledge graph}.

\begin{figure}[htbp]
  \centering
  \includegraphics[width=0.8\textwidth]{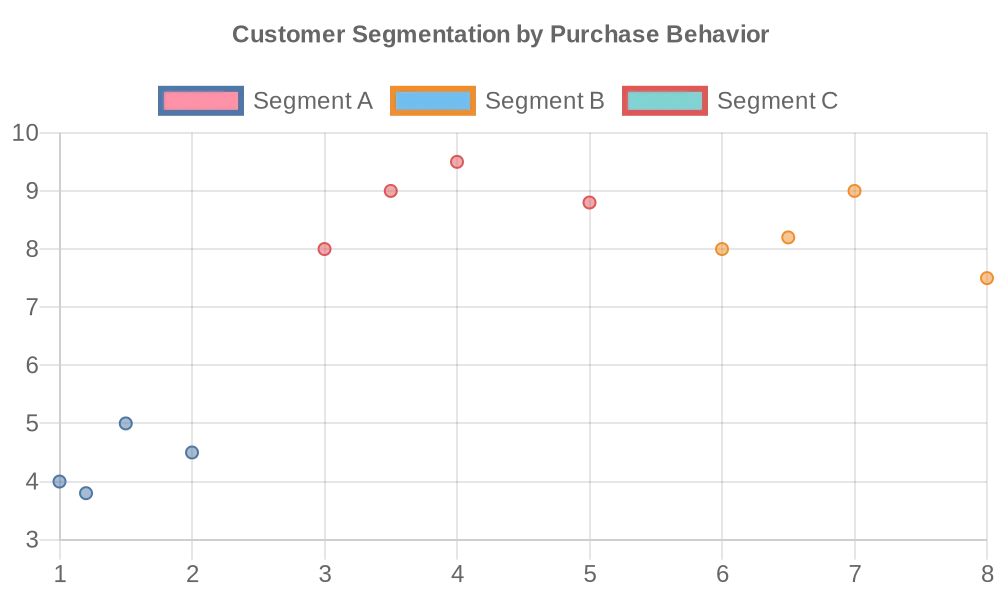}
  \caption{A simplified business knowledge graph. AI can analyze this structure to identify influential nodes (e.g., key customers), strong paths (e.g., effective sales funnels), and emerging clusters (e.g., new customer segments).}
  \label{fig:knowledge_graph}
\end{figure}

In this model, business entities (customers, products, campaigns) are represented as nodes, and their interactions are represented as edges. This visual framework allows entrepreneurs to literally see the patterns that the AI has uncovered:
\begin{itemize}
    \item \textbf{Tracking Trends:} By visualizing the graph over time, an SME can monitor how relationships are changing. For example, they might see the connection between a specific social media campaign and a new customer segment growing stronger, providing clear, visual confirmation of marketing ROI.
    \item \textbf{Root Cause Analysis:} When a KPI unexpectedly drops (e.g., sales for a specific product), a knowledge graph can help identify the root cause. The AI might highlight a weakening connection to a key supplier or a competing product that is attracting the same customers, providing an immediate, data-backed starting point for investigation.
    \item \textbf{Discovering Opportunities:} AI-powered graph analysis can uncover non-obvious relationships. It might discover that customers who buy Product A are highly likely to also buy Product C, even if they seem unrelated. This insight can be immediately translated into a targeted cross-selling campaign.
\end{itemize}

By using AI to not only make predictions but also to build and maintain a visual map of their business ecosystem, SMEs can transform abstract data into an intuitive, strategic dashboard. This graphical approach makes AI insights more accessible, actionable, and transparent, fostering a deeper, data-driven understanding of the business's core relationships and trends.

\section{Measuring the Return: Quantifying the Impact of AI on the Bottom Line}

While the functional benefits of AI are clear, the decision to invest ultimately hinges on its impact on the bottom line. The data shows that for SMEs, AI is not just an operational enhancement but a powerful driver of revenue growth, cost savings, and productivity.

\subsection{The Revenue Revolution}

The most compelling metric for any business initiative is its effect on revenue. For SMEs adopting AI, the results are overwhelmingly positive. A landmark survey found that \textbf{91\% of SMBs with AI say it directly boosts their revenue} \cite{salesforce2024smb}. This top-line growth is driven by enhanced capabilities across the sales and marketing funnel. AI-powered tools have been shown to increase qualified leads by up to 50\% and improve overall marketing efficiency by a similar margin \cite{explodingtopics2025, vereus2024}. Furthermore, a McKinsey study found that 63\% of businesses that implemented AI with the primary goal of reducing costs and optimizing resources also witnessed an unexpected boost in revenue, indicating that operational efficiency often translates directly into new growth opportunities \cite{makebot2025, mckinsey2019}.

\subsection{The Cost-Saving Imperative}

AI's ability to automate processes and optimize resource allocation delivers substantial cost reductions. Research indicates that SMEs can save \textbf{up to 30\% on overall operational costs} by strategically leveraging AI-powered expense optimization \cite{superagi2024}. These savings are realized across various business functions:
\begin{itemize}
    \item \textbf{Customer Support:} Implementing AI chatbots can cut customer support costs by as much as one-third by handling routine queries and reducing the need for a large human support team \cite{indatalabs2024}.
    \item \textbf{Procurement:} Automated contract analysis and supplier negotiation tools can yield savings of 10-15\% \cite{superagi2024}.
    \item \textbf{Labor:} Predictive scheduling and workforce optimization can reduce labor costs by 8-12\% \cite{superagi2024}.
\end{itemize}

For many small businesses, these percentages translate into tangible monthly savings, with surveys reporting that entrepreneurs are saving between \textbf{\$500 and \$2,000 per month} after adopting AI tools \cite{colorwhistle2025}.

\subsection{The Productivity Dividend}

Beyond direct financial metrics, AI delivers a significant "productivity dividend" by giving back the most valuable resource: time. Many SMEs report saving \textbf{20 or more hours per month} by automating repetitive and administrative tasks \cite{colorwhistle2025}. This reclaimed time allows entrepreneurs and their teams to focus on strategic, high-value activities like innovation, customer relationships, and business development. Specific case studies have quantified these gains, showing that AI can reduce the time spent on administrative tasks by 60\% and cut the time required for drafting meeting minutes by a factor of four \cite{activdev2025ops, vereus2024}. On a larger scale, companies like EchoStar have projected that their AI applications will save an astounding 35,000 work hours annually while boosting overall productivity by at least 25\% \cite{microsoft2025idc}.

\subsection{Table 1: AI ROI for SMEs - A Snapshot of Key Metrics and Case Studies}

The following table summarizes the quantifiable returns on investment that SMEs are achieving through AI adoption across various business functions. It provides a powerful, at-a-glance overview of the technology's impact, linking specific applications to measurable outcomes.

\begin{table}[htbp]
  \caption{AI ROI for SMEs - A Snapshot of Key Metrics and Case Studies}
  \label{tab:table1}
  \centering
  \begin{tabular}{lllll}
    \toprule
    \textbf{Business Function} & \textbf{AI Application} & \textbf{Key Metric} & \textbf{Quantified Result} & \textbf{Case Study / Source} \\
    \midrule
    Sales/Customer Service & AI Sales Assistant on Website & Qualified Meetings & +40\% in 3 months & \cite{activdev2025sales} \\
    Marketing & Churn Prediction (RFM Analysis) & Customer Churn & -15\% in 6 months & \cite{activdev2025churn} \\
    Marketing & Churn Prediction (RFM Analysis) & Customer Lifetime Value & +10\% & \cite{activdev2025churn} \\
    Human Resources & Automated Employee Onboarding & Time Saved per Hire & 2-3 hours & \cite{activdev2025hr} \\
    Operations & Automated Meeting Summaries & Time Spent on Minutes & Divided by 4 & \cite{activdev2025ops} \\
    Sales & AI-Powered Sales Outreach & Qualified Leads & +35\% in 3 months & \cite{vereus2024} \\
    Operations & General Workflow Automation & Operational Costs & -25\% in 6 months & \cite{vereus2024} \\
    Overall Business & General AI Tool Adoption & Monthly Cost Savings & \$500 - \$2,000 / month & \cite{colorwhistle2025} \\
    Overall Business & General AI Tool Adoption & Monthly Time Savings & 20+ hours / month & \cite{colorwhistle2025} \\
    Overall Business & AI Adopters vs. Non-Adopters & Revenue Boost & 91\% of adopters report increase & \cite{salesforce2024smb} \\
    \bottomrule
  \end{tabular}
\end{table}

\section{From Concept to Reality: A Strategic Roadmap for AI Adoption}

Successfully integrating AI requires a structured, strategic approach. Simply purchasing a tool is not a strategy. This section outlines a phased roadmap designed specifically for SMEs, guiding them from initial assessment to scalable, long-term value creation.

\subsection{Phase 1: Readiness Assessment \& Strategic Alignment (Weeks 1-4)}

Before adopting any AI technology, a thorough internal assessment is crucial. This foundational phase ensures that AI initiatives are aligned with business needs and grounded in reality.

\begin{itemize}
    \item \textbf{Define Business Goals:} The process must begin with business pain points, not technology solutions. Identify the most critical challenges or opportunities in your organization. Is it slow customer response times, inefficient inventory management, or an inability to generate qualified leads? Clearly defining the problem is the most important step \cite{behindthedesign2025, neurond2024}.
    \item \textbf{Data Audit:} AI models are entirely dependent on the data they are trained on. It is essential to assess the quality, quantity, and accessibility of your business data. Research shows that poor data quality is a primary barrier to successful AI implementation for SMEs, and only a small fraction of enterprises have data that is truly ready for AI ingestion \cite{makebot2025, arxiv2025maturity}.
    \item \textbf{Skills \& Culture Assessment:} Honestly evaluate your team's current understanding of AI and their readiness for new, data-driven workflows. Studies consistently identify knowledge gaps as the most significant obstacle for SMEs, so acknowledging this upfront is key to planning for necessary training and change management \cite{arxiv2025maturity, omdena2025}.
    \item \textbf{Tech Infrastructure Review:} Determine if your current technology stack—your CRM, accounting software, and other systems—can integrate with modern AI tools. For many SMEs, leveraging cloud-based AI services can mitigate the need for large, costly on-premise hardware investments \cite{aws2025, arxiv2025maturity}.
\end{itemize}

\subsection{Phase 2: Identify Quick Wins \& Pilot Projects (Weeks 5-8)}

The goal of this phase is to build momentum and demonstrate tangible value quickly, which is critical for securing buy-in and justifying further investment.

\begin{itemize}
    \item \textbf{Prioritize High-Impact, Low-Complexity Projects:} Avoid the temptation to tackle your most complex problem first. Instead, identify a "quick win"—a project that can deliver measurable value with minimal investment and risk. Examples include implementing a customer service chatbot to handle basic FAQs or using a generative AI tool to accelerate social media content creation \cite{jdmeier2024}.
    \item \textbf{Embrace Rapid Experimentation:} The accessibility of AI has lowered the barrier to entry. Leverage free trials or low-cost subscription models of AI tools to test concepts and validate their potential without significant upfront financial commitment \cite{synergy2024, makebot2025}.
    \item \textbf{Define Success Metrics (KPIs):} Establish clear, measurable Key Performance Indicators (KPIs) for your pilot project before it begins. Vague goals lead to ambiguous results. Be specific: "reduce average customer email response time by 30\%" or "increase marketing email click-through rates by 15\%" \cite{spaceo2025, vereus2024}.
\end{itemize}

\subsection{Phase 3: Implementation \& Integration (Weeks 9-16)}

With a validated pilot project, the next step is formal implementation and integration into existing workflows.

\begin{itemize}
    \item \textbf{Choose the Right Tools \& Partners:} Select AI solutions that are user-friendly, scalable, and cost-effective. Crucially, ensure they can integrate smoothly with your existing software stack to avoid creating new data silos \cite{neurond2024}. For more complex projects, consider partnering with a specialized AI agency. This can reduce implementation time by up to 60\% and cut costs by 30-40\% compared to in-house efforts \cite{vereus2024}.
    \item \textbf{Data Preparation \& Cleaning:} This is a non-negotiable step. The data identified in Phase 1 must be cleaned, structured, and formatted correctly for the chosen AI tool. This often represents a significant portion of the work in any AI project but is essential for accurate results \cite{synergy2024}.
    \item \textbf{Train Your Team:} AI adoption is as much about people as it is about technology. Provide comprehensive training to employees who will use the new tools. Communicate clearly how AI will augment their roles and make their jobs more effective, which is key to managing resistance to change and fostering adoption \cite{activdev2024}.
\end{itemize}

\subsection{Phase 4: Scaling \& Fostering a Data-Driven Culture (Ongoing)}

A successful pilot is not the end goal; it is the beginning. The true value of AI is unlocked when it is scaled across the organization and embedded in its culture.

\begin{itemize}
    \item \textbf{Analyze Pilot Results \& Iterate:} Use the data and feedback from your pilot project to analyze its performance against the KPIs defined in Phase 2. Use these learnings to refine the solution and your implementation process before attempting a broader rollout \cite{synergy2024}.
    \item \textbf{Develop a Long-Term Roadmap:} Based on the success of initial projects, build a strategic, long-term AI roadmap. This plan should prioritize future AI initiatives that align with core business objectives and build upon each other \cite{jdmeier2024, spaceo2025}.
    \item \textbf{Foster a Data-Driven Culture:} The ultimate goal is to transform the organization's mindset. This involves making data literacy a core competency for all employees, encouraging experimentation, publicly celebrating data-driven successes, and investing in tools that make data accessible to everyone, not just a few analysts \cite{velosio2025, confluent2024, googlecloud2024culture}.
\end{itemize}

Many SMEs successfully execute a pilot project but fail to scale the benefits because they treat AI as a series of disconnected tools. The most successful adopters will pursue a "Pilot-to-Platform" strategy. The ROI from an initial quick win, like a chatbot, should be used not just to fund the next pilot, but to invest in building a unified data infrastructure. This creates a central, clean, and accessible data platform that makes every subsequent AI deployment faster, cheaper, and more powerful, creating a compounding return on investment that transforms the entire enterprise.

\section{Navigating the Pitfalls: Managing Risks and Ethical Considerations}

While the potential of AI is immense, its adoption is not without challenges and risks. A proactive and responsible approach is essential for SMEs to navigate these pitfalls, build trust with stakeholders, and ensure the long-term sustainability of their AI initiatives.

\subsection{The Challenge of Implementation}

SMEs face a distinct set of barriers when implementing AI, which must be acknowledged and addressed in any adoption strategy.

\begin{itemize}
    \item \textbf{Cost \& ROI Uncertainty:} Despite the increasing affordability of AI tools, the initial implementation costs, including software, infrastructure, and potential consulting fees, can be a significant hurdle for businesses with limited capital \cite{mdpi2025, esade2025}.
    \item \textbf{Knowledge \& Skills Gap:} The most frequently cited barrier to AI adoption in SMEs is a lack of in-house expertise. Many business leaders and their teams lack a deep understanding of AI's capabilities and how to effectively implement it, creating a major obstacle to getting started \cite{arxiv2025maturity, omdena2025, oecd2024}.
    \item \textbf{Data Quality \& Availability:} Unlike large corporations with vast data repositories, SMEs often have smaller, less structured, and lower-quality datasets. Since the performance of AI models is directly dependent on data, this can be a significant technical challenge \cite{arxiv2025maturity}.
\end{itemize}

\subsection{Building a Framework for Responsible AI}

Implementing AI ethically is not just a matter of compliance; it is a business necessity that builds customer trust and mitigates significant risks. For SMEs, adopting a strong ethical framework can become a powerful competitive differentiator. In an era of growing consumer skepticism, SMEs that are transparent and responsible in their use of AI can build deeper customer loyalty than larger, more opaque competitors.

\begin{itemize}
    \item \textbf{Data Privacy \& Security:} Protecting sensitive customer and business data is paramount. When using third-party AI tools, especially generative AI platforms, it is critical to understand their data usage policies and to avoid inputting proprietary or personally identifiable information \cite{makebot2025, sba2025, iapp2024}. Compliance with data protection regulations like the GDPR is non-negotiable and must be a central component of any AI strategy \cite{smartosc2024, activdev2024}.
    \item \textbf{Algorithmic Bias:} AI systems learn from historical data, and if that data contains existing societal biases (e.g., in past hiring decisions or marketing imagery), the AI will learn and potentially amplify those biases. This can lead to discriminatory outcomes in areas like recruitment, credit assessment, and advertising \cite{sap2024a, triplel2024}. To mitigate this, SMEs must strive to use diverse and representative training data and regularly audit the outputs of their AI systems for fairness \cite{secureprivacy2025}.
    \item \textbf{Transparency \& Human Oversight:} AI should be deployed as a tool to augment human intelligence, not replace it entirely. It is crucial to maintain a "human-in-the-loop" for critical or sensitive decisions. All AI-generated content intended for external use, such as marketing materials or customer communications, should be reviewed by a person to ensure accuracy, appropriateness, and alignment with the brand's voice \cite{mckinsey2025, secureprivacy2025}.
    \item \textbf{Accountability:} Clear lines of responsibility for the outcomes of AI systems must be established within the organization. If an AI-driven decision leads to a negative outcome, there should be a clear process for addressing the issue and a designated individual or team accountable for its resolution \cite{secureprivacy2025, profiletree2024}. Proactively creating and publicizing an ethical AI charter can turn this potential risk into a brand asset, signaling to customers a commitment to responsible innovation.
\end{itemize}

\section{Technical Appendix: Getting Hands-On with AI}

This section provides practical, entry-level code examples to demystify the process of working with AI and machine learning. These tutorials are designed for beginners and can be run by an entrepreneur with some technical curiosity or a member of their team.

\subsection{Tutorial: Analyzing Customer Sentiment with Python \& a Cloud API}

\textbf{Business Goal:} To automatically analyze customer feedback from online reviews to quickly identify happy and unhappy customers, enabling rapid response to issues.

\textbf{Approach:} This tutorial uses Google Cloud's Natural Language API. This approach is ideal for SMEs as it leverages a powerful, pre-trained model, eliminating the need for complex model development and allowing for immediate results.

\textbf{Steps:}
\begin{enumerate}
    \item \textbf{Introduction to Sentiment Analysis:} Sentiment analysis tools inspect text to determine the prevailing emotional opinion, typically classifying it as positive, negative, or neutral. The API returns a `score` (from -1.0 for negative to 1.0 for positive) and a `magnitude` (the overall strength of the emotion) \cite{googlecloud2024sentiment, azure2024sentiment}.

    \item \textbf{Prerequisites:}
    \begin{itemize}
        \item A Google Cloud account with the Natural Language API enabled.
        \item Python 3 installed on your computer.
        \item The Google Cloud Client Library for Python installed (`pip install --upgrade google-cloud-language`).
        \item Authentication set up as described in the Google Cloud documentation \cite{googlecloud2024sentiment}.
    \end{itemize}

    \item \textbf{Code Walkthrough:} The following Python script sends a sample review to the API and prints the sentiment score.

\begin{lstlisting}[language=Python, caption={Analyzing Customer Reviews with Google Cloud API}]
# Import the necessary Google Cloud client library
from google.cloud import language_v1

def analyze_sentiment(text_content):
    """
    Analyzes the sentiment of a block of text.
    """
    # Instantiate a client
    client = language_v1.LanguageServiceClient()

    # The text to analyze
    document = language_v1.Document(content=text_content, type_=language_v1.Document.Type.PLAIN_TEXT)

    # Detects the sentiment of the text
    response = client.analyze_sentiment(document=document)
    sentiment = response.document_sentiment

    print(f"Text: {text_content}")
    print(f"Sentiment Score: {sentiment.score:.2f}")
    print(f"Sentiment Magnitude: {sentiment.magnitude:.2f}")

    return sentiment

# --- Business Application ---
# A list of customer reviews for your product
customer_reviews = [
    "The product is amazing! I absolutely love the quality and design.",
    "Completely disappointed. The item arrived broken and customer service was unhelpful.",
    "It's an okay product. Does the job, but nothing special."
]

print("--- Analyzing Customer Reviews ---")
for review in customer_reviews:
    sentiment = analyze_sentiment(review)
    # Flag negative reviews for immediate follow-up
    if sentiment.score < 0:
        print("!! NEGATIVE REVIEW DETECTED - ACTION REQUIRED!!")
    print("-" * 20)
\end{lstlisting}

    \item \textbf{Business Application:} The script loops through a list of reviews. By checking if the `sentiment.score` is less than zero, it can automatically flag negative feedback. An SME could adapt this code to read reviews from a file, a database, or a social media feed, creating an automated system to alert the customer service team to unhappy customers in real-time.
\end{enumerate}

\subsection{Tutorial: Building a Simple Predictive Model with Python \& Scikit-learn}

\textbf{Business Goal:} To build a simple classification model that predicts whether a customer will churn (stop using a service) based on their usage patterns.

\textbf{Approach:} This tutorial uses `scikit-learn`, a popular and user-friendly Python library for machine learning. It demonstrates the fundamental workflow of training a model on historical data and using it to make predictions on new data.

\textbf{Steps:}
\begin{enumerate}
    \item \textbf{Introduction to Classification:} A classification model predicts which category an item belongs to. In this case, we want to classify customers into two categories: "Will Churn" or "Will Not Churn" \cite{sklearn2024, datacamp2026}.

    \item \textbf{Prerequisites:}
    \begin{itemize}
        \item Python 3 installed.
        \item The `scikit-learn` and `pandas` libraries installed (`pip install scikit-learn pandas`).
    \end{itemize}

    \item \textbf{Data Loading \& Preparation:} We will use a sample dataset representing customer data. In a real-world scenario, this would come from your CRM or sales database.

\begin{lstlisting}[language=Python, caption={Loading and Preparing Data for Churn Prediction}]
import pandas as pd
from sklearn.model_selection import train_test_split
from sklearn.tree import DecisionTreeClassifier
from sklearn.metrics import accuracy_score, classification_report

# Create a sample DataFrame representing customer data
# In a real project, you would load this from a CSV file: df = pd.read_csv('customer_data.csv')
data = {'monthly_usage_hours': [10, 45, 5, 50, 2, 25, 30, 40, 15, 8],
        'support_tickets_filed': [1, 5, 0, 4, 0, 2, 3, 4, 1, 0],
        'churned': [0, 1, 0, 1, 0, 0, 1, 1, 0, 0]} # 1 = Churned, 0 = Not Churned
df = pd.DataFrame(data)

# Separate the features (X) from the target variable (y)
X = df[['monthly_usage_hours', 'support_tickets_filed']]
y = df['churned']
\end{lstlisting}

    \item \textbf{Train-Test Split:} It is crucial to split the data into a training set and a testing set. The model learns from the training set, and its performance is evaluated on the unseen testing set. This prevents the model from simply memorizing the data and ensures it can generalize to new customers \cite{sklearn2024, datacamp2026}.

\begin{lstlisting}[language=Python, caption={Splitting Data for Training and Testing}]
# Split data into 80% for training and 20% for testing
X_train, X_test, y_train, y_test = train_test_split(X, y, test_size=0.2, random_state=42)
\end{lstlisting}

    \item \textbf{Model Training:} We will train a `DecisionTreeClassifier`, a simple yet intuitive model that makes predictions by learning a series of `if-then-else` rules from the data.

\begin{lstlisting}[language=Python, caption={Training the Decision Tree Model}]
# Initialize the model
model = DecisionTreeClassifier(random_state=42)

# Train the model on the training data
model.fit(X_train, y_train)
print("Model training complete.")
\end{lstlisting}

    \item \textbf{Evaluation \& Prediction:} Now, we use the trained model to make predictions on the test data and evaluate its accuracy.

\begin{lstlisting}[language=Python, caption={Evaluating the Model and Making Predictions}]
# Make predictions on the test data
y_pred = model.predict(X_test)

# Evaluate the model's performance
accuracy = accuracy_score(y_test, y_pred)
print(f"\nModel Accuracy: {accuracy:.2f}")
print("\nClassification Report:")
print(classification_report(y_test, y_pred))

# --- Business Application: Predict churn for new customers ---
new_customers = pd.DataFrame({
    'monthly_usage_hours': [55, 12],
    'support_tickets_filed': [6, 1]
})

new_predictions = model.predict(new_customers)
print("\n--- Predictions for New Customers ---")
for i, prediction in enumerate(new_predictions):
    status = "Will Churn" if prediction == 1 else "Will Not Churn"
    print(f"New Customer {i+1} (Usage: {new_customers['monthly_usage_hours'][i]}, Tickets: {new_customers['support_tickets_filed'][i]}): {status}")
\end{lstlisting}
This simple model demonstrates how an SME can leverage its own data to build predictive capabilities, moving from reactive problem-solving to proactive, data-driven strategy.
\end{enumerate}

\section{Conclusion: The Future-Ready SME}

The evidence presented throughout this report converges on a single, unequivocal conclusion: Artificial Intelligence is a transformative force that is now squarely within the reach of Small and Medium-sized Enterprises. The question for entrepreneurs and business leaders is no longer \textit{if} they should adopt AI, but \textit{how quickly and strategically} they can integrate it into the very core of their operations. The most significant barriers to this transformation are not technological or financial, but are instead rooted in a lack of strategic understanding and a clear, actionable roadmap—a gap this report has sought to fill.

The ROI is proven, with the vast majority of SME adopters reporting direct boosts to revenue, substantial cost savings, and invaluable gains in productivity. The tools are more accessible and affordable than ever before, and a phased, pilot-driven approach can de-risk investment and build crucial organizational momentum. By focusing first on specific business pain points and leveraging AI as a force multiplier for lean teams, SMEs can achieve a level of operational efficiency and market intelligence previously reserved for their largest competitors.

As business leaders look to the future, the pace of change will only accelerate. The next frontier of AI will further empower agile businesses:
\begin{itemize}
    \item \textbf{AI Agents:} The evolution from simple chatbots to autonomous AI agents is already underway. These agents will be capable of performing complex tasks on behalf of employees—proactively scheduling meetings, managing sales outreach, and even ordering supplies—further augmenting the capabilities of small teams \cite{microsoft2024trends, googlecloud2025trends}.
    \item \textbf{Hyperautomation:} This strategic approach involves combining AI, machine learning, and robotic process automation to automate as many business processes as possible. The goal is to create a highly efficient, self-optimizing organization where human talent is focused almost exclusively on high-value, strategic work \cite{sap2024b, ibm2024hyper, pagerduty2024}.
    \item \textbf{Democratization of AI:} The rise of no-code and low-code AI platforms will continue to lower the barrier to entry, empowering non-technical employees to build and deploy custom AI solutions to solve their own departmental challenges, fostering a culture of bottom-up innovation \cite{usc4am2024}.
\end{itemize}

The SMEs that thrive in the coming decade will be those that view AI not as a series of disparate tools, but as a foundational platform for growth. They will prioritize building a data-driven culture, invest in unifying their data infrastructure, and commit to responsible, ethical implementation as a source of competitive advantage. These future-ready enterprises will not just survive the AI revolution; they will lead it, defining the future of their industries through unparalleled agility, intelligence, and customer value.

\bibliographystyle{unsrtnat}

\end{document}